\documentclass[aps,prd,preprint,tightenlines,superscriptaddress]{revtex4}
\usepackage{epsfig}
\usepackage{graphicx}
\usepackage{psfrag}
\usepackage{amsmath,amssymb}
\usepackage{colordvi}
\usepackage{color}
\usepackage{amsfonts}
\usepackage{enumerate}
\usepackage{slashed,bm}

\begin{document}

\title{Parton Physics on Euclidean Lattice}
\author{Xiangdong Ji}
\affiliation{INPAC, Department of Physics and Astronomy, Shanghai Jiao Tong University, Shanghai, 200240, P. R. China}
\affiliation{Maryland Center for Fundamental Physics, Department of Physics,  University of Maryland, College Park, Maryland 20742, USA}
\date{\today}
\vspace{0.5in}
\begin{abstract}
I show that the parton physics related to correlations of quarks and gluons on the light-cone
can be studied through the matrix elements of frame-dependent, equal-time correlators in
the large momentum limit. This observation allows practical calculations of parton properties
on an Euclidean lattice. As an example, I demonstrate how to recover the leading-twist
quark distribution by boosting an equal-time correlator to a large momentum.

\end{abstract}

\maketitle

Since deep-inelastic scattering experiments at the Stanford Linear Accelerator Center
in late 1960's, the proton structure has been probed in various hard
scattering processes~\cite{Gao:2013xoa}. The results are the quark and gluon (parton) momentum
distributions and correlations, as well as distribution amplitudes, etc. The parton
physics involves intrinsically light-cone correlations in the sense that
all quark and gluons fields are separated along the light-cone, defined with the real
Minkowski time $t$. The information is extremely useful for
understanding the non-perturbative structure of the proton and, for example,
for calculating the Higgs production cross section at Large Hadron Collider~\cite{science}.

Computing the parton physics from the fundamental theory, quantum chromodynamics (QCD),
has been difficult. In principle, a light-cone correlation is best calculated using
the proton's wave function in light-front coordinates~\cite{brodsky}. However, despite many years
of effort, a realistic light-front proton wave function has not yet been established.
On the other hand, in the formulation of non-perturbative QCD on a Euclidean lattice, one cannot directly calculate
time-dependent correlations. Instead, one can only compute moments of parton distributions and distribution amplitudes,
which are matrix elements of local operators. However, the difficult
grows considerably for higher moments for technical reasons~\cite{negele}.

In this paper, I present a direct approach to compute parton physics on a Euclidean lattice
through Lorentz boost. To demonstrate
this, let us consider the quark distribution in a proton,
\begin{eqnarray}
   q(x, \mu^2) &=& \int \frac{d\xi^-}{4\pi} e^{-ix\xi^-P^+}  \langle P|\overline{\psi}(\xi^-)
   \gamma^+ \\
   && \times \exp\left(-ig\int^{\xi^-}_0 d\eta^- A^+(\eta^-) \right)\psi(0) |P\rangle \nonumber \ ,
\end{eqnarray}
where the nucleon momentum $P^\mu$ is along the $z$-direction, $P^\mu = (P^0, 0, 0, P^z)$, and $\xi^\pm
=(t \pm z)/\sqrt{2}$ is the light-cone coordinates with $t$ as the physical time, $\mu^2$ is the renormalization scale, $A^+$
is a gluon potential matrix in the fundamental representation, $\psi$ is a quark field of flavor $q$. The
above expression is boost-invariant along $z$, i.e., independent $P^z$; in particular it is valid in the rest frame
where $P^z=0$.

To find a way to calculate $q(x,\mu^2)$ on a Euclidean lattice, it is useful to
exhibit the origin of the light-cone correlation: The parton distribution has also been formulated in terms of
the local twist-2 operators, which are defined as
\begin{equation}
     O^{\mu_1...\mu_n}= {\overline \psi}\gamma^{(\mu_1}iD^{\mu_2} ... iD^{\mu_n)} \psi
                 - {\rm trace} \ ,
\end{equation}
where $(\mu_1...\mu_2)$ indicates that all the indices are symmetrized, the trace terms include
operators with at least one factor of the metric tensor $g^{\mu_i \mu_j}$ multiplied by
operators of dimension $(n+2)$ with $n-2$ Lorentz indices, etc. Its matrix elements in the proton state
are,
\begin{equation}
    \langle P|O^{\mu_1...\mu_n}(\mu^2)|P\rangle =
    2a_n(\mu^2) (P^{\mu_1} ... P^{\mu_n} - {\rm trace}) \ ,
\end{equation}
and the parton distribution is related to the local matrix elements through $\int dx x^{n-1} q(x, \mu^2)
= a_n(\mu^2)$ with even $n$. The time-dependent correlation for the parton distribution
in Eq. (1) is recovered by taking all the
components as $+$ in Eq. (3),
\begin{equation}
   \langle P|O^{+...+}(\mu^2)|P\rangle =2 a_n(\mu^2) P^+...P^+ \ .
\end{equation}
Thus, the light-cone correlation is {\it kinematically} connected with
+ components of the nucleon four-momentum.

To eliminate the time dependence, we consider the matrix elements of the twist-two operator with $\mu_1=\mu_2=...=\mu_n=z$,
and in the nucleon state with large $P^z$.  Then
\begin{equation}
   O^{z...z} = {\overline \psi}\gamma^{z}iD^{z} ... iD^{z} \psi - {\rm trace} \ ,
\end{equation}
where again the trace terms contain operators with at most $n-2$ $z$'s. According
to Lorentz symmetry, the matrix elements of the trace terms are at most
$(P^z)^{n-2}$ times $\Lambda^2_{\rm QCD}$. Similarly the right hand side of Eq. (3) is
$2a_n(\mu^2)\left[(P^z)^n-\lambda M^2(P^z)^{n-2}-...\right]$, where $\lambda$ is a numerical number of
order 1, $M$ is the nucleon mass, and ... represents terms
with still lesser powers of $P^z$. Thus, we conclude that
\begin{eqnarray}
     && \langle P| {\overline \psi}\gamma^{z}iD^{z} ... iD^{z} \psi|P\rangle
       = 2a_n(\mu^2) (P^z)^n\times \left[ 1 \right.\\ && \left.+ {\cal O}\left(\Lambda^2_{\rm QCD}/(P^z)^2,  M^2/(P^z)^2\right)\right] \nonumber
\end{eqnarray}
where terms other than the first are power-suppressed in the large $P^z$ limit.

Inverting the above result in term of a time-independent, non-local expression, we find
in the large $P^z$ limit,
\begin{eqnarray}
   q(x,\mu^2, P^z) &=& \int \frac{dz}{4\pi} e^{izk^z}  \langle P|\overline{\psi}(z)
   \gamma^z \\
   && \times \exp\left(-ig\int^{z}_0 dz' A^z(z') \right)\psi(0) |P\rangle \nonumber
 \\ && + {\cal O}\left(\Lambda^2/(P^z)^2,  M^2/(P^z)^2\right) \ ,  \nonumber
\end{eqnarray}
where $x= k^z/P^z$.
An intuitive way to understand the above result is to consider
the Lorentz transformation of a line segment connecting $(0,0,0,z)$ with the origin of the coordinates.
As the boost velocity approaches the speed of light, the space-like line segment is tilted to the
light-cone direction. Of course, it cannot literally be on the light-cone because the invariant length
cannot change for any amount of boost. However, this slight off-light-cone-ness only introduces power corrections
which vanish asymptotically.

Intuitively, the parton distribution in Eq. (1) can be regarded for the nucleon
in the rest frame but the probe is traveling at the speed of light, and hence the light-cone correlation;
On the other hand, the distribution  Eq. (7) is obtained from a static probe on a nucleon traveling at
the speed of light. This late quantity is related to deep-inelastic scattering in a frame where the virtual photon
momentum contains only the space component $q^\mu = (0, Q)$, and the nucleon momentum is $P^z \sim Q/2x$.
One can also obtain a similar expression for the quark helicity distribution with the replacement of
$\gamma^z$ by $\gamma^z\gamma_5$. As for the gluon distribution, we have,
\begin{eqnarray}
g(x,\mu^2, P_z) = \int \frac{dz}{4\pi k^z}
e^{izk^z} \langle P|F^{3\mu}(z)  \\
\times \exp\left(-ig \int^z_0 dz' A^z(z')\right)
F^{~3}_\mu(0)|P\rangle \nonumber
\end{eqnarray}
in the large $P^z$ limit, where $A^z$ is a matrix in adjoint representation and $i$ sums over the transverse directions.
And similarly, we have the polarized gluon distribution,
\begin{eqnarray}
\Delta g(x,\mu^2, P_z) = i\int \frac{dz}{4\pi k^z}
e^{izk^z} \langle P|F^{3\mu}(z)  \\
\times \exp\left(-ig \int^z_0 dz' A^z(z')\right)
\tilde F^{~3}_\mu(0)|P\rangle \nonumber \ ,
\end{eqnarray}
from which one can integrate to get $\Delta G$ ($\tilde F^{\mu\nu} = 1/2 \epsilon^{\mu\nu\alpha\beta}F_{\alpha\beta}$). However, we also find another way to calculate
$\Delta G$ on lattice without explicitly doing this integral~\cite{Ji:2013fga}.

\begin{figure}[hbt]
\begin{center}
\includegraphics[width=30mm]{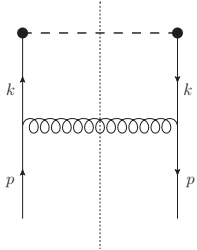}
\caption{One-loop correction to the parton distribution in Eq. (7) in $A^z=0$ gauge.}
\label{phqcorrection}
\end{center}
\end{figure}

Taking the infinite-momentum limit is a subtle process in field theories, and the ultraviolet (UV) divergences
must be taken care of properly in the loop integrals with a correct limiting procedure.
In our case, we need to take the limit after the renormalization procedure, whereas
the standard light-cone result in Eq. (1) has $P^z\rightarrow \infty$ taken before the
renormalization. Thus in modern language, the light-cone distribution is an effective field theory
of the quantity in Eq. (7), in which there are large logarithms $\ln P^z$ in
perturbation theory, which can be transformed into the standard renormalization
scale dependence in the light-cone distribution through matching conditions~\cite{jizhangzhao}. One can see the interplay of
the different limits by computing the one-loop correction in the physical $A^z=0$ gauge (see Fig. 1),
\begin{eqnarray}
    q(x, \mu^2, P^z) = \frac{\alpha_s}{2\pi}T(x) ~\frac{\Lambda}{P^z}+    \frac{\alpha_s}{2\pi} P(x) ~ \ln\frac{\Lambda P^z}{\lambda^2}
\end{eqnarray}
where $\Lambda = \sqrt{\mu^2 + (1-x)^2(P^{z})^2}-(1-x)P^z$, $\lambda$ is an infrared mass cut-off and
$\mu^2$ is an UV cut-off (renormalization scale);
$T(x) =C_F x/(1-x)^2$, and $P(x) = C_F (1+x^2)/(1-x)$ with $C_F=4/3$ is the standard Altarelli-Parisi kernel~\cite{AP}. In the limit of
$P^z\rightarrow \infty$ first, $\Lambda \sim \mu^2/P^z$, the first term is power-suppressed and the second term is the
standard AP result. However, in the limit of $\mu^2\rightarrow \infty$ first, which is of our interest here,
the first term is linearly divergent, whereas the second term becomes $\ln (\mu P^z/\lambda^2)$. The connection
of the two limits is reflected through the following factorization theorem up to power corrections,
\begin{equation}
    q(x, \mu^2, P^z)  = \int^1_x \frac{dy}{y} Z\left(\frac{x}{y},\frac{\mu}{P^z}\right) q(y, \mu^2)
\end{equation}
where $Z$ has a perturbative expansion in $\alpha_s$,
\begin{equation}
             Z(x, \mu/P^z) = \delta (x-1) + \frac{\alpha_s}{2\pi} Z^{(1)}(x, \mu/P^z) + ...
\end{equation}
with $Z^{(1)} = T(x)(\mu/P^z) + P(x)\ln (P^z/\mu)$. Taking $P^z \rightarrow \infty$ first
yields $Z(x, \mu/P^z) = \delta(x-1)$. Thus the large logarithmic dependence on $P^z$ in $q(x,\mu^2,P^z)$ can be transformed into the
renormaliazation scale dependence through the above matching condition.
On lattice, the matching must be recalculated up to a constant accuracy using the standard approach~\cite{ren}.

Therefore, Eq. (11) can be used to calculate the parton distribution $q(x,\mu^2)$ on lattice by measuring
the time-independent, non-local quark correlator $\overline{\psi}(z)\gamma^zL(z,0)\psi(0)$ with
$L(z,z_0) = \exp\left(-ig\int^{z}_{z_0} dz' A^z(z') \right)$ in a state with increasingly large $P^z$
(maximum $\sim 1/a$ with $a$ denoting lattice spacing). 
Because the smallest and the largest momenta on lattice is $1/L_z a$ and $1/a$, respectively,
the smallest $x$ that one can resolve is $1/L_z$, with $L_z$ the number of lattice sites in the $z$-direction.
Thus to reach $x = 10^{-2}$, one has to have $~100$ lattice sites. On the other hand, the present approach can get a
reasonable description of large $x$ partons, emphasized by higher moments of the distribution. Thus, the method is
in a sense complementary to the traditional moment approach. On the other hand, for a fixed $x$, large $P^z$ implies large $k^z$,
and hence small $z$. Thus, a fast nucleon experiences a Lorentz contraction in the $z$-direction which requires
increasing resolution (anistropic lattices such as ones used for spectroscopy calculations~\cite{Edwards:2011jj}) to study its longitudinal structure.

The above approach can be generalized to any light-cone correlations in hadron physics.
A partial list of interesting physical quantities includes:
\begin{itemize}
\item{Generalized parton distributions (GPD's)~\cite{Ji:1996ek}, for which one has the same twist-two operators, but the external
states are off-diagonal. Thus, for example, the GPD's $E(x,\xi, t)$ and $H(x, \xi, t)$  can be calculated
from the following matrix elements,
\begin{eqnarray}
  F(x, \xi, t, P^z) &=& \int \frac{dz}{2\pi} e^{-izk^z}\langle P'|\overline\psi(-z/2)\gamma^z \\ && \times L(-z/2,z/2)\psi(z/2)|P\rangle \nonumber
\end{eqnarray}
with $\vec{P}'+\vec{P}=2(0,0,\overline{P}^z)$, in the $P^z\rightarrow \infty$, $\xi= (P^{z'}-P^z)/\overline{P}^z$, $x=k^z/\overline{P}^z$-fixed limit. Similarly one can obtain other twist-two GPD's with different insertions of the Dirac matrix and nucleon polarization states.
Thus one can now explore the three-dimensional dependence on $x$, $\xi$ and $t$ in full detail without model assumptions~\cite{Hagler:2003jd}.}

\item{Transverse-momentum dependent (TMD) parton distribution~\cite{collins}.  For the Dirac matrix $\gamma^z$, for example, we can obtain the following TMD,
\begin{eqnarray}
  && q(x,k_\perp, P_z, \mu^2)\nonumber \\ &=& \int \frac{dz}{4\pi} d^2\vec{r}_\perp e^{i(zk^z+\vec{k}_\perp \vec{r}_\perp)}  \langle P|\overline{\psi}(\vec{r}_\perp,z)
    \\ && \times L^\dagger\left(\pm\infty; (\vec{r}_\perp,z)\right)\gamma^z L(\pm\infty; 0)\psi(0) |P\rangle \nonumber
\end{eqnarray}
where $P^z$ must be large and $L\left(\pm\infty; (\vec{r}_\perp, z)\right) = e^{-ig\int^{\pm\infty}_{z} dz' A^z(\vec{r}_\perp, z')}$. The dependence
on $P^z$ now contains large logarithms $\ln P^z$ and does not disappear in the large $P^z$ limit, and therefore is shown in the function explicitly.
One can derive an evolution equation for $P^z$~\cite{collins}.
The gauge links here extend from the location of the quark field to $\pm \infty$
alone the $z$-direction only, and therefore they involve only the $A^z$ component of the gauge potential. Here $\pm\infty$ corresponds to two different kinematic
conditions in which the distributions are probed. In gauge choices where gauge potential vanishes at infinity, the above expression is gauge invariant.
However, for a finite lattice with all gauges averaged over, the two gauge links must be connected by another gauge link $L((\vec{r}_\perp, \pm\infty);(0,\pm \infty))$ in the transverse-space direction. Similar expressions work for TMDs defined with other Dirac structure. An exploratory study of $x$-moments of TMD distributions has been made on the lattice where gauge links are taken as staple in spatial directions~\cite{Musch:2010ka}. The connection to light-cone kinematics
is made through invariant decompositions, not through boosts, although the distributions themselves are defined at large $P^z$.}

\item{Wigner distributions~\cite{viewing}, which are generating functions for both GPD's and TMD's.  The definition follows
 from a TMD operator sandwiched in a non-forward nucleon states. Thus one can calculate them in the large $P^z$ limit as, for example,
\begin{eqnarray}
  &&  W(x,k_\perp,b_\perp, P^z, \mu^2) \\ &&
  = \int \frac{dz}{4\pi} d^2\Delta_\perp d^2\vec{r}_\perp e^{-i\Delta_\perp \cdot b_\perp} e^{i(zk^z+\vec{k}_\perp \vec{r}_\perp)}\nonumber
    \\ && \times \langle P'|\overline{\psi}(\vec{r}_\perp,z)L^\dagger(\pm\infty; (\vec{r}_\perp,z))\gamma^z L(\pm\infty; 0)\psi(0) |P\rangle \nonumber
\end{eqnarray}
where $\Delta_\perp =(P'-P)_\perp$ and again a transverse gauge link is needed to ensure gauge invariance on a finite lattice. We also
note that in the large $P^z$ limit, the $P^z$ dependence does not disappear.}

\item{Light-cone amplitudes, which are the matrix elements between the hadron states and the QCD vacuum,
of the light-cone correlations of the quark and gluon interpolating fields. We can now calculate them as
the matrix element of spatial correlation in the large momentum limit. For example, the leading light-cone
amplitude of the $\pi^+$ meson recovers as the large $P^z$ limit of
\begin{equation}
\phi_0(x, P^z) 2P^z = \int \frac{dz}{2\pi} e^{izk^z}
\langle 0|\overline{d}(0)\gamma^z\gamma_5
L(0,z)u(z)| \pi^+(P)\rangle \ .
\end{equation}
These light-cone amplitudes can be probed from the hard exclusive processes such as the
form factors~\cite{Lepage:1980fj,Belitsky:2002kj}.
There has been much discussion about the $x$-dependence of $\phi_0(x)$ amplitude in the literature.}

\item{Light-cone wave functions. In the light-front coordinates, the nucleon state can be
written as an infinite sum of Fock states with wave-function amplitudes~\cite{brodsky}.
In a series of publications~\cite{Ji:2002xn}, we have systematically classified
the wave-function amplitudes for pion and proton up to four partons, and connected them with
the matrix elements of quark and gluon correlations on the light-cone and transverse directions.
All these are gauge-invariant when light-cone gauge links are inserted from the locations of the fields
to the infinity along the light-cone direction. These amplitudes again can be calculated on lattice as
the infinite-momentum limit of the similar correlation functions with fields connected by gauge links
going to infinity along the $z$-direction. One important application is to the exclusive decays
of B mesons where many factorization formulas have been derived with light-cone wave functions
as a part of the decay amplitudes~\cite{li}.}

\item{Higher-twist parton distributions. The complete list of twist-two, -three and -four parton
distributions in unpolarized, longitudinally-polarized and transversely-polarized nucleon has been
worked out in the literature~\cite{Ellis:1982cd}. They all contains quark and gluons fields separated
along the light-cone. In the spirit of this paper, they can be obtained from related spatial correlation
functions with gauge links along the $z$-direction in the limit of $P^z \rightarrow \infty$. In particular, the
jet quenching parameter $\hat q$ might be calculated this way~\cite{Majumder:2012sh}.}

\item{...}

\end{itemize}
It will be interesting to see these quantities can now be explored in lattice QCD calculations.

In conclusion, I have shown that the light-cone correlations of quarks and gluons
can be calculated by boosting the matrix elements
of spatial correlations to a large momentum. This is particularly
useful for lattice QCD calculations of these experimental observables
which have been very difficult to tackle from the first principle in the past.
To be sure, studying a large momentum hadron on lattice is computationally
still challenging, but at least this could be achieved when computational power continues
improving.

I thank J. W. Chen, M. Engelhardt, C. Liu, K. F. Liu, Y. Jia, Z. B. Kang, Q. Wang, Z. T. Wei, and Y. Zhao for useful
comments. I particularly appreciate H.-N. Li and J. W. Qiu for their comments about the IR behavior and
factorization of the Euclidean distribution. This work was partially supported by the U.
S. Department of Energy via grants DE-FG02-93ER-40762, and a grant (No. 11DZ2260700) from the Office of
Science and Technology in Shanghai Municipal Government, and by National Science Foundation of China (No. 11175114).


\begin{thebibliography}{99}

\bibitem{Gao:2013xoa}
See for example,
  J.~Gao, M.~Guzzi, J.~Huston, H.~-L.~Lai, Z.~Li, P.~Nadolsky, J.~Pumplin and D.~Stump {\it et al.},
  arXiv:1302.6246 [hep-ph];
  H.~-L.~Lai, M.~Guzzi, J.~Huston, Z.~Li, P.~M.~Nadolsky, J.~Pumplin, C.~-P.~Yuan and ,
  Phys.\ Rev.\ D {\bf 82}, 074024 (2010)
  [arXiv:1007.2241 [hep-ph]].

\bibitem{science}
The CMS Collaboration, Science,  December Issue 1569-1575, 2012;
The ATLAS Collaboration, Science, December Issue 1576-1582, 2012

\bibitem{brodsky}
  S.~J.~Brodsky,
  In *Adelaide 1999, Lightcone QCD and nonperturbative hadron physics* 15-26
  [hep-ph/0004211].

\bibitem{negele}
See for example, 
  J.~W.~Negele,
  Nucl.\ Phys.\ A {\bf 711}, 281 (2002)
  [hep-lat/0211022];

  P. Hagler {\it et al.}  [LHPC Collaboration],
  Phys.\ Rev.\ D {\bf 77}, 094502 (2008)
  [arXiv:0705.4295 [hep-lat]].


\bibitem{AP}
  G.~Altarelli and G.~Parisi,
  Nucl.\ Phys.\ B {\bf 126}, 298 (1977).


\bibitem{Ji:2013fga}
  X.~Ji, J.~-H.~Zhang and Y.~Zhao,
  arXiv:1304.6708 [hep-ph].


\bibitem{Edwards:2011jj}
  R.~G.~Edwards, J.~J.~Dudek, D.~G.~Richards and S.~J.~Wallace,
  Phys.\ Rev.\ D {\bf 84}, 074508 (2011)
  [arXiv:1104.5152 [hep-ph]].
  
  
\bibitem{ren}
  S.~Capitani,
  Nucl.\ Phys.\ B {\bf 597}, 313 (2001)
  [hep-lat/0009018]; 
  S.~Capitani,
  Nucl.\ Phys.\ B {\bf 592}, 183 (2001)
  [hep-lat/0005008].
  S.~Capitani,
  Phys.\ Rept.\  {\bf 382}, 113 (2003)
  [hep-lat/0211036].

\bibitem{Ji:1996ek}
  X.~Ji,
  Phys.\ Rev.\ Lett.\  {\bf 78}, 610 (1997)
  [hep-ph/9603249];
  D.~Mueller, D.~Robaschik, B.~Geyer, F.~M.~Dittes, and J.~Horejsi,
  Fortsch.\ Phys.\  {\bf 42}, 101 (1994)
  [hep-ph/9812448].
  X.~Ji,
  Ann.\ Rev.\ Nucl.\ Part.\ Sci.\  {\bf 54}, 413 (2004).

\bibitem{Hagler:2003jd}
  P.~Hagler {\it et al.}  [LHPC and SESAM Collaborations],
  Phys.\ Rev.\ D {\bf 68}, 034505 (2003)
  [hep-lat/0304018];
  P.~.Hagler {\it et al.}  [LHPC Collaboration],
  Phys.\ Rev.\ D {\bf 77}, 094502 (2008)
  [arXiv:0705.4295 [hep-lat]].

\bibitem{collins}
  J.~C.~Collins, and D.~E.~Soper,
  Nucl.\ Phys.\ B {\bf 194}, 445 (1982).
  J.~C.~Collins, and D.~E.~Soper,
  Nucl.\ Phys.\ B {\bf 193}, 381 (1981)
  [Erratum-ibid.\ B {\bf 213}, 545 (1983)]
  [Nucl.\ Phys.\ B {\bf 213}, 545 (1983)].
  X.~Ji, J.~-p.~Ma, and F.~Yuan,
  Phys.\ Rev.\ D {\bf 71}, 034005 (2005)
  [hep-ph/0404183].


\bibitem{Musch:2010ka}
  B.~U.~Musch, P.~Hagler, J.~W.~Negele, and A.~Schafer,
  Phys.\ Rev.\ D {\bf 83}, 094507 (2011)
  [arXiv:1011.1213 [hep-lat]];
  B.~U.~Musch, P.~Hagler, M.~Engelhardt, J.~W.~Negele, and A.~Schafer,
  Phys.\ Rev.\ D {\bf 85}, 094510 (2012)
  [arXiv:1111.4249 [hep-lat]].

\bibitem{viewing}
  X.~Ji,
  Phys.\ Rev.\ Lett.\  {\bf 91}, 062001 (2003)
  [hep-ph/0304037].
  A.~V.~Belitsky, X.~Ji, and F.~Yuan,
  Phys.\ Rev.\ D {\bf 69}, 074014 (2004)
  [hep-ph/0307383].
  X.~Ji, X.~Xiong, and F.~Yuan,
  Phys.\ Rev.\ Lett.\  {\bf 109}, 152005 (2012)
  [arXiv:1202.2843 [hep-ph]].
  X.~Ji, X.~Xiong, and F.~Yuan,
  arXiv:1207.5221 [hep-ph].

\bibitem{Lepage:1980fj}
  G.~P.~Lepage, and S.~J.~Brodsky,
  Phys.\ Rev.\ D {\bf 22}, 2157 (1980);
  A.~V.~Efremov, and A.~V.~Radyushkin,
  Phys.\ Lett.\ B {\bf 94}, 245 (1980).

 \bibitem{Belitsky:2002kj}
  A.~V.~Belitsky, X.~Ji, and F.~Yuan,
  Phys.\ Rev.\ Lett.\  {\bf 91}, 092003 (2003)
  [hep-ph/0212351].

\bibitem{Ji:2002xn}
  M.~Burkardt, X.~Ji, and F.~Yuan,
  Phys.\ Lett.\ B {\bf 545}, 345 (2002)
  [hep-ph/0205272];
  X.~Ji, J.~-P.~Ma, and F.~Yuan,
  Nucl.\ Phys.\ B {\bf 652}, 383 (2003)
  [hep-ph/0210430]; 
  X.~Ji, J.~-P.~Ma, and F.~Yuan,
  Phys.\ Rev.\ Lett.\  {\bf 90}, 241601 (2003)
  [hep-ph/0301141].
  X.~Ji, J.~-P.~Ma, and F.~Yuan,
  Eur.\ Phys.\ J.\ C {\bf 33}, 75 (2004)
  [hep-ph/0304107].

\bibitem{li}
  Y.~Y.~Keum, H.~-N.~Li, and A.~I.~Sanda,
  Phys.\ Rev.\ D {\bf 63}, 054008 (2001)
  [hep-ph/0004173];
  Y.~-Y.~Keum, H.~-n.~Li, and A.~I.~Sanda,
  Phys.\ Lett.\ B {\bf 504}, 6 (2001)
  [hep-ph/0004004].
\bibitem{Ellis:1982cd}
  R.~K.~Ellis, W.~Furmanski, and R.~Petronzio,
  Nucl.\ Phys.\ B {\bf 212}, 29 (1983);
  J.~-w.~Qiu, and G.~F.~Sterman,
  Nucl.\ Phys.\ B {\bf 353}, 105 (1991);
  X.~Ji,
  Nucl.\ Phys.\ B {\bf 402}, 217 (1993);
  P.~Hoodbhoy, and X.~Ji,
  Phys.\ Rev.\ D {\bf 50}, 4429 (1994)
  [hep-ph/9307304].

\bibitem{Majumder:2012sh}
  A.~Majumder,
  arXiv:1202.5295 [nucl-th].
\end{thebibliography}
\end{document}